\documentstyle[12pt,aaspp4]{article}

\input epsf
\def\ni{\noindent}
\def\bar{\overline}
\def\bbE{\bar {\bf E}}
\def\bw{\bar{\omega}}
\def\beq{\begin{equation}}
\def\pineee{\end{equation}}
\def\lsim{\mathrel{\rlap{\lower4pt\hbox{\hskip1pt$\sim$}}
    \raise1pt\hbox{$<$}}}
\def\gsim{\mathrel{\rlap{\lower4pt\hbox{\hskip1pt$\sim$}}
    \raise1pt\hbox{$>$}}}

\def\bfA{{\bf A}}
\def\bfa{{\bf a}}
\def\bfe{{\bf e}}
\def\bfB{{\bf B}}

\def\bv{\bar V}

\def\ts{\times}
\def\lb{\langle}
\def\rb{\rangle}
\def\curl{\nabla {\ts}}

\def\bfv{{\bf v}}

\def\bfj{{\bf j}}

\def\bfe{{\bf e}}
\def\bfw{{\bomega}}

\def\bfb{{\bf b}}

\def\bfB{{\bf B}}

\def\bbB{\overline {\bf B}}

\newcommand\figstages      {1}
\newcommand\figfinalstate  {2}

\newcommand\tableism      {1}
\newcommand\tablegalaxy   {2}
\newcommand\tablesim      {3}
\newcommand\tablescales   {4}

\newcommand\ee{\end{equation}}

\newcommand\B{{\bf B}}
\newcommand\bb{{\bf b}}

\newcommand\vv{{\bf v}}
\newcommand\x{{\bf x}}

\newcommand\bnabla{\mbox{\boldmath $\nabla$}}

\begin{document}

\nonumber
\setcounter{equation}{0}

\centerline{\large\bf
 Effect of Fractional Kinetic Helicity}
\centerline{\large\bf on Turbulent Magnetic Dynamo Spectra}

\medskip

\author{Jason Maron\altaffilmark{1} and Eric G. Blackman\altaffilmark{2}}
\affil{1. Department of Physics, UCLA,  Los Angeles, CA}
\affil{2. Department of Physics \&Astronomy and Laboratory for
Laser Energetics, University of Rochester, Rochester NY 14627}

\centerline {(submitted to Ap. J. Lett.)} 

\begin{abstract}  



Magnetic field amplification in astrophysics ultimately
requires an understanding of magnetohydrodynamic turbulence.
Kinetic helicity has long been known to be important
for large scale field growth in forced MHD turbulence,
and has been recently demonstrated numerically to be 
asymptotically consistent with slow mean field dynamo action 
in a periodic box. 
Here we show numerically that 
the magnetic spectrum at and below the forcing scale 
is also strongly influenced by kinetic helicity. 
We identify a critical value, $f_{h,crit}$ 
above which the magnetic spectrum develops maxima at 
wavenumber $= 1$ scale {\it and} at the forcing scale,
For $f< f_{h,crit}$ the field peaks only at the resistive scale.
Kinetic helicity may thus be important
not only for generating a large scale
field, but also for establishing observed peaks
in magnetic spectra at the forcing scale.
The turbulent Galactic disk provides an example where 
both large scale ($>$ supernova forcing scale) fields
and small scale ($\le$ forcing scale, with peak at forcing scale) 
fields are observed.  We discuss this, and the potential application
to the protogalaxy, but also emphasize the limitations 
in applying our results to these systems.


\bigskip

{{\bf Subject Headings}:  MHD; turbulence; 
ISM: magnetic fields;  galaxies: magnetic fields; 
stars: magnetic fields; methods: numerical}

\end{abstract}

\vfill
\eject
       
\bigskip


\section{Introduction}

The origin of magnetic fields and 
the dynamics of 3-D magnetohydrodynamic (MHD) turbulence in astrophysical sources are problems of long standing interest 
(e.g. Cowling 1957; Moffatt 1978; Parker 1979; Krause \&  R\"adler 1980;
Zeldovich et al. 1984). 
The standard in situ mean field dynamo (MFD) model  (Moffatt 1978; Parker 1979;
Krause \&  R\"adler 1980)
of the large scale (= scales greater than the turbulent forcing) magnetic
field origin can be thought of as a framework
for understanding an inverse cascade of magnetic
helicity, initiated by a forcing of kinetic helicity (Pouquet et al. 1975). 
While the role of kinetic and magnetic helicities are 
important for in situ
non-local inverse cascade models of large scale fields, 
or MFDs, the 
small scale (= scales at or below the turbulent forcing
scale) dynamo does not explicitly require helicity to amplify total 
magnetic energy density (Zeldovich et al 1983; Parker 1979).
Non-helically forced turbulent amplification of the small scale fields,
and fully helical  forced growth of large scale fields
have been recently simulated (Cho \& Vishniac 2001; 
Chou 2001; Brandenburg 2001; Maron \& Cowley 2001).


But there is an important subtlety which has not yet been addressed.
Though numerical work generically shows 
that the total energy of the small scale 
field in turbulent media saturates to of order
 equipartition with the 
kinetic energy spectrum, non-helical small scale dynamos 
produce a peak of the magnetic energy spectrum
on the resistive scale for magnetic Prandtl number
$\ge 1$, not on the forcing scale (Chou 2001; Maron \& Cowley 2001).
This contradicts, for example, observations of the Galactic
magnetic field which has a peak in the spectrum on the
turbulent forcing scale (Beck et al. \& 1996) and maximally helical 
simulations (Brandenburg 2001).
Here we show that forcing with 
varying levels of  fractional kinetic helicity affects 
the overall spectral shape at large $and$ small scales.

In section 2 we discuss the equations and the simulations. 
In section 3 we give the results and in section 4 the interpretations.
We conclude in section 5.

\section{Equations and Numerical Scheme}

We investigate forced helical MHD turbulence. 
We write the magnetic field  in velocity
units and so define $\bb \equiv \B/ \sqrt{4 \pi}$,
where $\bfB$ is the magnetic field. 
Incompressibility is assumed throughout, so we set 
density $\rho = 1$, and $\nabla\cdot\bfv=0$, where $\bfv$ is velocity.
We include the thermal pressure $P$, and
the magnetic pressure in the total pressure 
$p=P+b^2/2$ and  assume isotropic kinetic and magnetic
viscosities, $\nu_v$.
The MHD equations become,
\begin{equation}
\partial_t \vv = - \vv \cdot \nabla \vv- \nabla p 
+\bb \cdot \bnabla \bb,
+ \nu_v \nabla^{2} \vv,
\label{eq:mhda}
\end{equation}
\begin{equation}
\partial_t\bb = - \vv \cdot \bnabla \bb + \bb \cdot \bnabla \vv
+ \nu_b \nabla^{2}\bb,
\label{eq:mhdb}
\end{equation}

\noindent To relate $p$ to $\vv$ and $\bb$, we take the divergence of equation
\ref{eq:mhda}, which upon inversion, yields
\beq
p=\int \frac{d^3x^\prime}{4\pi}
\frac{(\bnabla\vv:\bnabla\vv-\bnabla\bb:\bnabla\bb)}{|\x^\prime-\x|}.
\label{eq:Pvb}
\ee


A random forcing field with energy $\epsilon_f \cdot \Delta\!t$ is
generated at each time step and added to the existing velocity field,
where $\epsilon_f$ is the average forcing power. The amplitudes of the
forcing field Fourier modes are assigned according to a specified
power spectrum, with the energy selected from a Boltzmann distribution.
The mode phases are random within the constraint of
divergencelessness. 
We input kinetic helicity ${\bf v} \cdot \nabla\times {\bf v}$
at the forcing wavenumber of 
$s=k/2\pi =4.5$ by making a randomly determined
subset of the Fourier modes maximally helical, leaving the rest
unchanged. The fraction of maximally helical modes is $f_h$, which we
denote ``fractional helicity.''
In contrast, simulations of  
Maron \& Cowley (2001)  invoked zero mean magnetic field and 
zero mean kinetic helicity.  Only fractional random fluctuations 
of the kinetic helicity of order 10\% were present. 
The magnetic helicity was also initially zero and subsequently fluctuated 
about zero at an amplitude of 10\%  of the maximum.  
The equations of MHD are solved spectrally.
The turbulence is incompressible and the boundaries are periodic. 
(Including compressibility, while ultimately important for detailed
applications, is not expected to have a dramatic effect on the qualitative
conclusions herein regarding the role of helicity.)
Wave numbers and physical scales are related by
$\lambda k = 2 \pi$. 
Viscosity and resistivity are of the $k^2$ type
($\nu_v \nabla^2 v$ and $\nu_b \nabla^2 b$).
The code is exhaustively discussed in Maron \& Goldreich (2001).
We note that at each time step, the time 
derivative of the magnetic helicity is equal to the 
volume integral of the current helicity times the resisitivity.
The helicity conservation equation is satisfied and 
this is important in what follows.



The other key parameters are as follows:
the magnetic 
Prandtl number is $Pr = \nu_{v}/ \nu_b \sim \lambda_{\nu_v}^2 /\lambda_{\nu_b}^2$,
the ratio of the viscosity to magnetic diffusivity,
where  $\lambda_{\nu_v}$ and $ \lambda_{\nu_b}$, are the viscous 
and resistive scales respectively. 
We denote $v_\lambda$ and $b_\lambda$ as the speed and magnetic field
at scale $\lambda$, and $v_f$ and $\lambda_f$ as the forcing scale
RMS velocity and forcing scale respectively.
When $Pr\ge 1$, the scales have the ordering $\lambda_f > \lambda_{\nu_v}
\geq \lambda_{\nu_b}$.

\section{Results}

We show results here for a selection of $64^3$ simulations which
is sufficient to identify the basic effects of fractional
helicity on the location of energy peaks. 
The $f_h$ ranges from 
$0$ to $1$ by increments of $0.1$ for simulations
A0 through A10. For all A0-A10, we used a $64^3$ grid, $s_f=4.5$,
$\nu_v=3\times 10^{-3}$, $\nu_b=1\times 10^{-3}$, and $Pr=3$.

The usual kinetic and magnetic energy spectra are
defined as the quantities inside  the energy integrals 
$E_v = \int E_v(s) d\!s $ and $ E_b = \int E_b(s) d\!s $ respectively.
The spectra for a range of values of $f_h$ are shown in 
Fig 1.  The time evolution of the $f_h=1$ case
is shown in Fig 2 and the time growth of magnetic helicity
is shown in Fig 3. 
Notice in these figures that for $f_h \gsim f_{h,crit}\sim 0.5$
the peak at the forcing scale grows as does the large scale
field.  For $f_h< f_{h,crit}$, the large scale field decays, no peak
appears at the forcing scale, and the magnetic helicity in the box
grows very weakly, if at all.
Though we have presented only $Pr=3$ cases in the figures,
we have also performed simulations with $Pr=9$ 
and found that $f_{h,crit}\sim 0.7$. Thus we find that 
$f_{h,crit}$ increases with $Pr$. 

We checked for hysteresis by
using the saturated state of the $f_h=1$ simulation of Fig. 3
as the initial condition for
another simulation with $f_h=0.4$. 
We found that the magnetic helicity subsequently decayed
to the same value as in the simulation which started with a weak
mean field with initial $f_h=0.4$. There was no evidence for hysteresis.

\vspace{-.1cm} \hbox to \hsize{ \hfill \epsfxsize8cm
\epsffile
{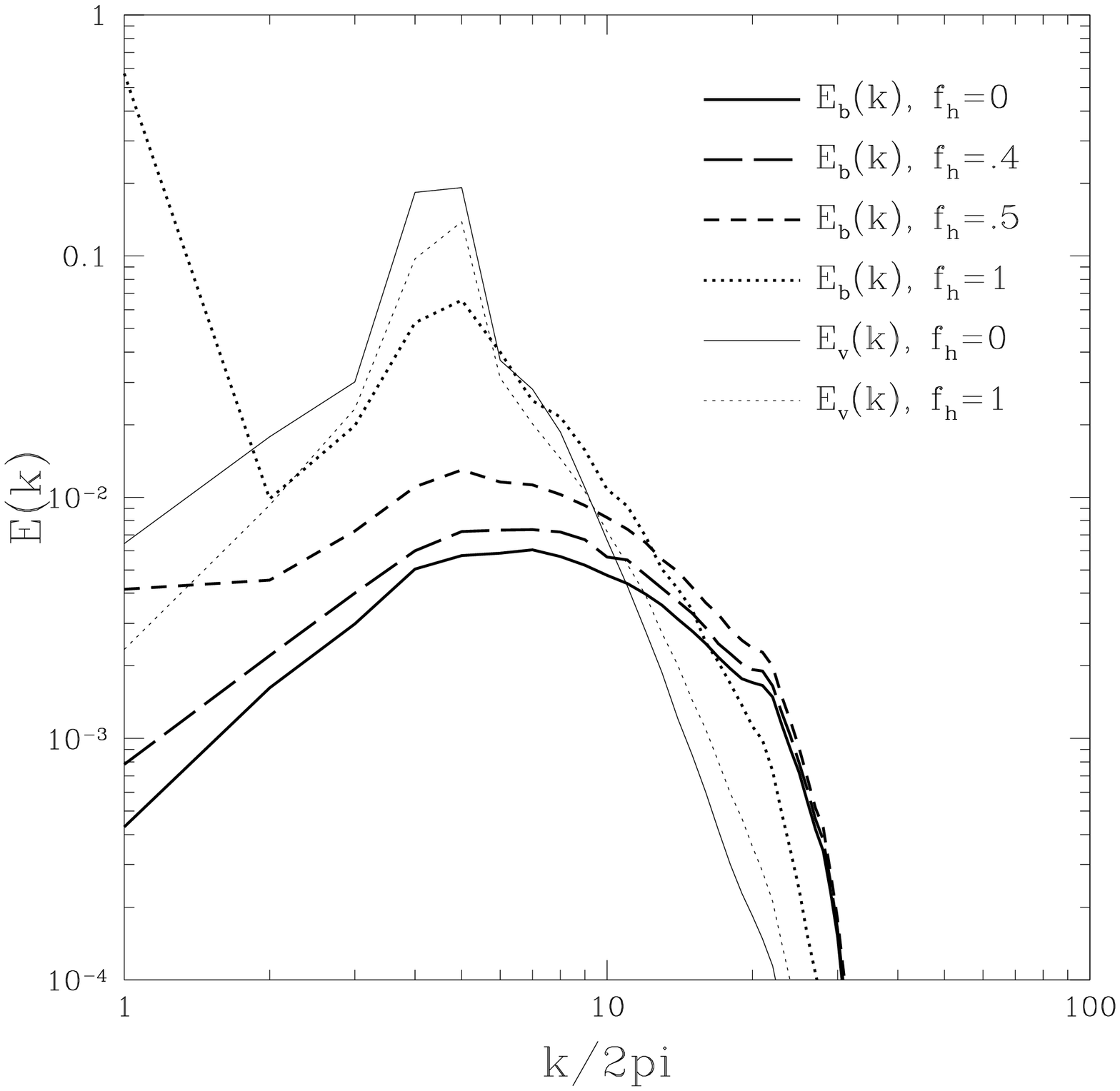} \hfill } \noindent {\it Figure 1:
Saturated kinetic and magnetic energy spectra for successive values of
$f_h$. The simulations are A0, A4, A5, and A10. The kinetic spectra
are identical for $f_h \le 0.4$}

\vspace{-.1cm} \hbox to \hsize{ \hfill \epsfxsize8cm
\epsffile{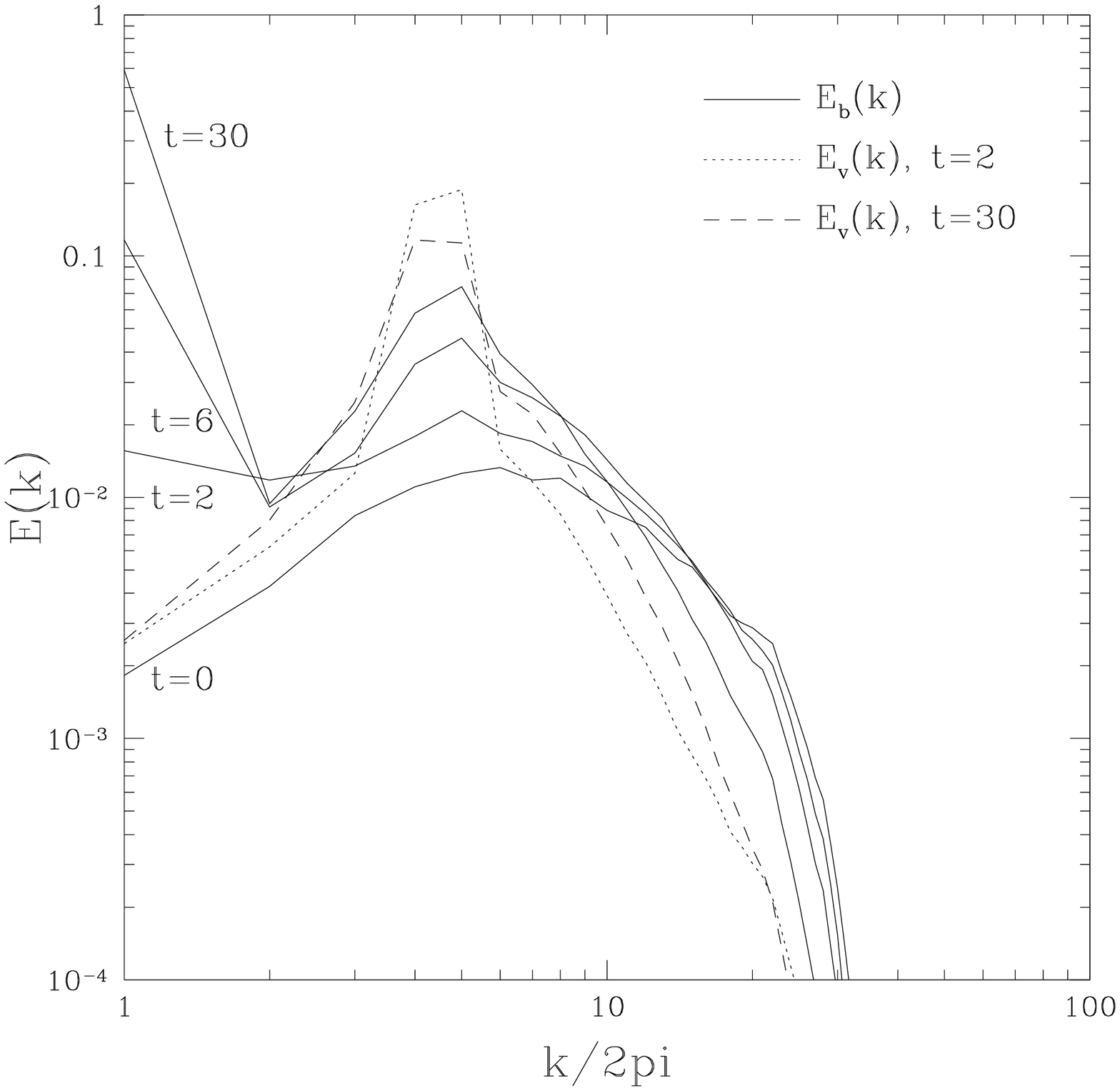} \hfill } \noindent {\it Figure 2:
Time sequence of kinetic and magnetic energy spectra for simulation A10.}

\vspace{-.1cm} \hbox to \hsize{ \hfill \epsfxsize8cm
\epsffile{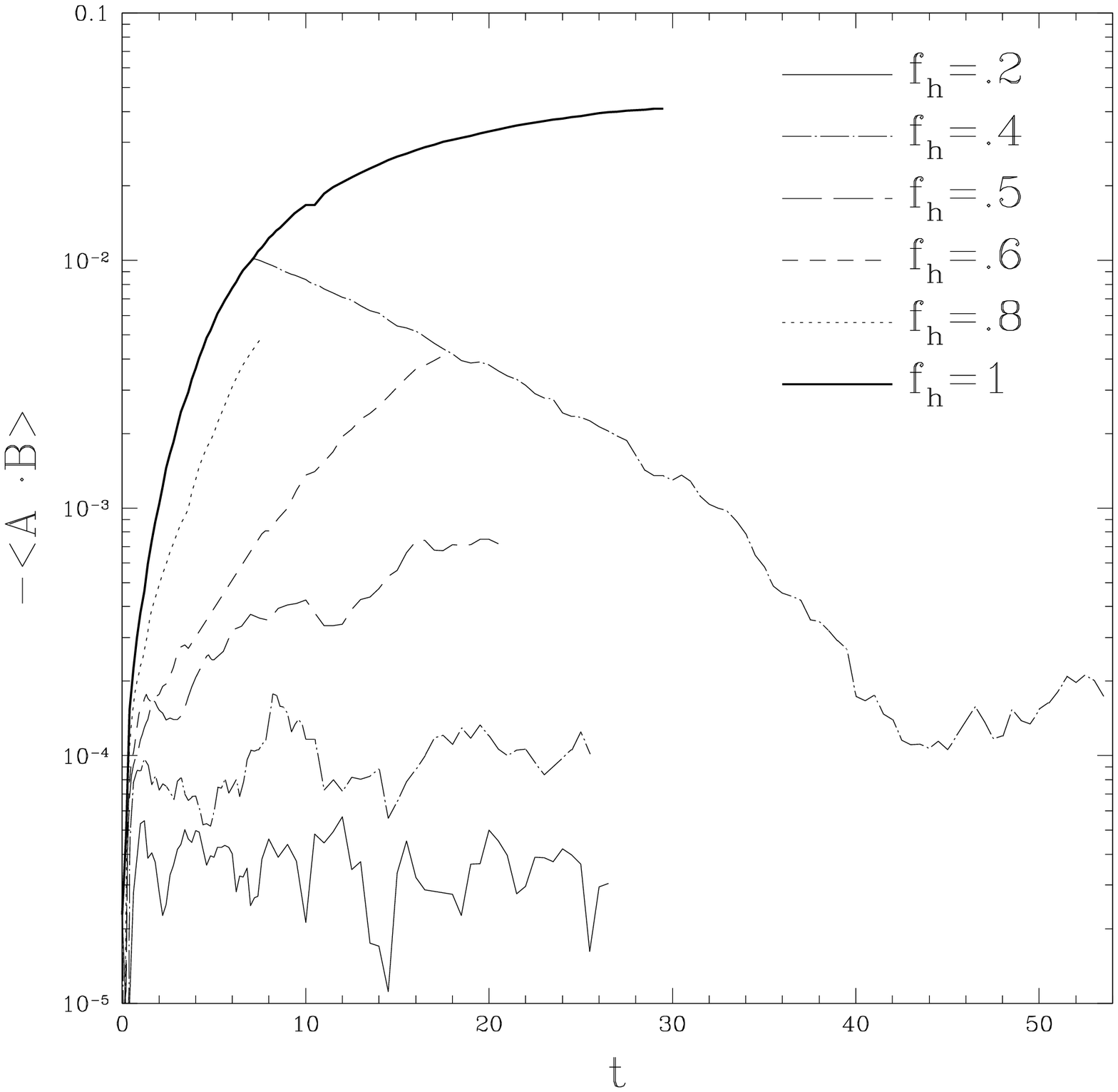} \hfill } \noindent {\it Figure 3: Evolution of
the magnetic helicity as a function of $f_h$ for  simulations from
series A. At $t=7$, simulation A10 ($f_h=1$) has developed a large
scale magnetic field.}

\section{ Discussion}

Forced helical turbulence was studied
systematically in Pouquet et al. (1976) 
using the Eddy Damped Quasi-Normal Markovian (EDQNM)
closure scheme (Orsag 1970). Forcing 
with kinetic helicity leads to segregation of magnetic helicity 
because the magnetic helicity growth equation
has a source term that depends on the kinetic helicity.
The kinetic helicity inputs one sign of magnetic helicity
at small scales but the opposite sign is generated
on large scales.  The growth of the large scale field, 
represents a non-local inverse cascade of the magnetic helicity
from the forcing scale and can be interpreted 
as an $\alpha^2$ MFD (Pouquet et al. 1976; Brandenburg 2001; 
Blackman \& Field 2001b; Field \& Blackman 2002; also Verma 2001).
Averaged over a periodic box, 
the time evolution of the total magnetic helicity 
satisfies $\partial_t\lb\bfA \cdot \bfB\rb=-2\nu_b \lb {\bf J}\cdot{\bf B}\rb$,
where $\bf A$ is the vector potential and $\bf J$ is the current density.
If we divide $\lb\bfA \cdot\bfB\rb$ into small scale and large scale
contributions, we can see that after the large scale
helical field energy grows to $k_{s}B_{s}^2/k_{l}$,
where $s,l$ refer to the dominant small and large scale,
the large scale magnetic helicity dominates (Fig. 3).
The growth saturates as long 
as there remains a net current helicity, but 
the growth rate is resistively limited, 
implying a ``slow'' 
(decreases with increasing $k_{\nu_b}$) MFD.  
The large scale field growth of Fig.
3 for all $f_h \ge f_{h,crit}$ is slow in this sense.
(Note however that there is a short initial ``fast'' phase.  
Field \& Blackman (2002) and Blackman (2002),
discuss this in the context of a dynamical quenching model
which fits Brandenburg (2001) and the results herein, and the asymptotic
quenching of Gruzinov \& Diamond 1994; Bhattacharjee \& Yuan 1995).

The sign of the magnetic helicity of the growing
large scale field  is opposite to
that of the kinetic helicity. This is  consistent with 
MFD theory if the kinetic helicity dominates 
the $\alpha$ effect of the MFD:
A positive kinetic helicity means that $\alpha$ would be
negative.  But the growth of the magnetic helicity associated
with the large scale field is $\propto \alpha {B_l}^2$ 
(Brandenburg 2001, Blackman \& Field 2001; Field \& Blackman 2002)
so that a positive input kinetic helicity, which gives a negative
$\alpha$,  produces a negative large scale magnetic helicity. 
Fig. 3  shows that
$\lb {\bf A}\cdot {\bf B}\rb$ is 
dominated by the large scale contribution.

That the large scale field growth proceeds asymptotically ``slow'',  
would seem to threaten the relevance for e.g. the Galaxy
as it is commonly argued that the MFD
of the Galaxy has to be ``fast.'' 
(c.f. Ruzmaikin et al 1988; Zweibel \& Heiles 1997).
Unlike a periodic box, real astrophysical rotators have boundaries, 
shear, and have helicity driven by the combination of
the underlying rotation and stratification with a spatial variation
in transport coefficients.
So these differences will ultimately need to be studied before 
results from periodic box solutions like ours can be directly 
applied to astrophysical large scale field growth (Blackman \& Field 2001).
Also, the large scale field for $f_h> f_{h,crit}$ in the simulation 
becomes super-equipartition as seen in Figs. 1 \& 2,
because it is nearly force-free. 
That being said, the enterprise of periodic box simulations
is very worthwhile; even comparing the difference between
periodic boxes with more realistic simulations will be
important to our further understanding MHD turbulence and 
allows the demonstration of some principles that emerge
in the simplest possible forced non-linear dynamo system.

In spite of these caveats, the results of the present simulations are 
provocative: the same $f_{h,crit}$ determines both whether 
the large scale field grows $and$ 
whether a peak grows at the forcing scale. 
The large scale field growth and the presence of the small scale
peak at the forcing scale appear to be 
intimately related. Magnetic helicity undoubtedly also plays a role
drain of the small scale peak from the resistive 
scales for $f_{h} > f_{h,crit}$.
Consider the role of magnetic helicity conservation 
in mode interactions via a slightly more general
version of the argument of (Frisch et al. 1975):
Suppose an initial state of maximal magnetic helicity is 
confined to wave numbers $k_m$ and $k_n$, with $k_m<k_n$
and suppose the magnetic field
dominates the energy at these wavenumbers.  We then have  
$E_b(k_m)+E_b(k_n)=E_T (k_p)$
and
$
H_b(k_m)+H_b(k_n)= E_b(k_m)/k_m + E_b(k_n)/k_n \le   E_T(k_p)/k_p
$, where $E_T$ means the total energy.
The last inequality can only be satisfied if 
$k_p < k_n$.  
This argument applies only 
when the scales $k_m$ and $k_n$ are magnetically dominated,
though $k_p$ need not be (a point not addressed in Frisch et al. 1975).
An initial state for which the field is dominant,
and therefore satisfies the validity
criterion, is the small scale saturated state shown in Fig. 1  for $f_h=0$.
We can reason that when sufficient magnetic helicity is imposed, 
it, and its associated energy would drain from the small scales, at least  
until the field reaches equipartition with the velocity.
This is qualitatively consistent with the observed 
deficit in the magnetic energy from the 
large $k$ region in figures 1 and 2 for $f_h> f_{h,crit}$, as 
compared to $f_h < f_{h,crit}$.

The growth of the actual peak in magnetic energy at the forcing scale
is also aided by the fact that a forced kinetic helicity
reduces the non-linear transfer term in the 
Navier-Stokes equation.  The non-linear term 
$-\bfv\cdot\nabla\bfv = \bfv\ts{\bf \omega}-\nabla v^2$.
When helicity is present, the $\bfv\ts{\bf \omega}$ term
is reduced.  For sub-sonic turbulence, 
the $v^2$ contribution to the evolution equation should be
inconsequential.  Thus, since the main cascade driver 
is reduced for helical turbulence, the kinetic 
energy requires more time to cascade, providing a bit
more time for this energy to be transfered directly into 
stretching the magnetic field near the input scale.
Though a cascade of magnetic energy steadily
drains the field from the input scale, the hold up of the kinetic energy
cascade means that there is more time to resupply the field
to a larger amplitude before draining, compared to the $f_h < f_{h,crit}$
case.

The total kinetic energy density for 
$k\ge k_{f}$ is fixed and is always
larger than the total magnetic energy density for $k\ge k_f$.
If $f_h<f_{h,crit}$, then 
there is a significant non-helical part of the field which feels
no tendency to inverse cascade. This fraction
piles up quickly on the small scales (Chou 2001; Maron \& Cowley 2001). 
If this fraction dominates,
then the spectrum will be dominated by the non-helical turbulence
dynamics.  If $f_h>f_{h,crit}$, the magnetic
energy associated with the helicity, which inverse cascades, dominates.
A lower limit on $f_{h,crit}$ can be found from 
modeling the large scale field growth 
as an $\alpha^2$ dynamo, which has a growth rate of $\alpha-\beta s_1$,
where $\beta$ is turbulent diffusion and $s_1=1$ is the growing large scale
wave number.  But initially $\alpha$ and $\beta$ have their
kinematic values and $\alpha/s_1\beta 
\sim 2f_h s_f/(3s_1)$,
and so initial growth requires at least  
$f_h > s_1/s_f\sim 0.33 = f_{crit}$.  This 
is roughly consistent with our results within small factors of order 1.

At risk of applying our ``slow'' dynamo 
spectral shape results too cavalierly, we consider
the implications for the Galaxy.
The Galactic field has both a large scale ($\gsim 2$kpc) 
and a small scale component ($\lsim 100$pc) (e.g. Zweibel \& Heiles 1997).
The small scale field (which has magnitude $\sim$ few times larger than 
the large scale field) appears to have a peak at the forcing scale,
as does the kinetic energy (Armstrong et al. 1995), 
and the two are in near equipartition with $v\sim b\sim 10$km/s.
Ignoring the ``slow'' vs. ``fast'' issue for the moment, 
our results  
would imply that the Galactic magnetic  spectrum (with its small scale
peak at the forcing scale) is only consistent with turbulence 
forced with $f_h > f_{h,crit}$. 
Note that we have been considering only externally forced turbulence,
as opposed to self-generated turbulence from shear.
This particular assumption, at least, is 
consistent with the Galaxy, where supernovae are the primary
driver (Sellwood \& Balbus 1999).  
(Given that our box dynamo is ``slow'' we should note that 
well motivated analytic interpretations for generation of ``fast'' 
large scale magnetic energy 
in sheared rotators which appeal less explicitly to kinetic
helicity, or not at all have been studied (see Balbus \& Hawley 1998;
Vishniac \& Cho 2000), as do specific proposals for 
how boundaries might enable fast helical $\alpha-\Omega$ dynamos (Blackman 2002). We do not discuss these further but note
that in the Galaxy, a net magnetic flux in the Galactic disk in addition
to magnetic energy seems to be needed.)

Our results might also suggest that kinetic helicity would play a role in the 
protogalactic small scale dynamo model (Kulsrud et al. 1997) 
of the large scale Galactic field. 
In this model, the large scale field of the Galaxy 
results from gravitational collapse and flux freezing of the
small scale protogalactic field.
The model requires that small scale dynamos generate
significant power at the forcing scale, and our results would suggest
this is only possible in $Pr\ge 1$ plasmas when $f_h \ge f_{h,crit}$.

\section{Conclusions}
 
The magnetic spectrum of 
MHD turbulence forced in a periodic box with fractional 
kinetic helicity above a critical value $f_{h,crit}$,  
saturates with two peaks: a large scale peak and
a peak at the forcing scale, when $Pr\ge 1$. 
If $f_h < f_{h,crit}$ there 
is only one peak in the spectrum, and it is at the resistive scale.
Though the turbulence is not strictly Alfv\'enic for any $f_h$, 
it is much more nearly  Alfv\'enic for larger $f_h$. 
The range of scales, the boundary conditions, and the actual
nature of helical forcing pose obstacles 
when comparing non-sheared periodic box simulations 
with real astrophysical rotators. 
Nevertheless, the important qualitative implication of our results 
is that helical forcing not only influences the large scale magnetic
field spectra of forced turbulent systems
but may also help account for  observed peaks at the forcing scale.


\ni 
EB acknowledges DOE grant DE-FG02-00ER54600. JM benefitted from 
supercomputers at the National Center for Supercomputing Applications
at UIUC (1000 CPU hours), and from their very helpful staff.
We thank S. Boldyrev, A. Brandenburg, S. Cowley, G. Field, 
J. McWilliams, V. Pariev, A. Schekochihin, and E. Vishniac 
for discussions.

\bigskip

\centerline{\bf References}

\def\item{\ni}

\ni 
Armstrong J.W., Rickett B.J., and 
Spangler S.R.,  ApJ, 1995, {443} 209 

\ni Balbus S.A \& Hawley J., 1998, Rev. Mod. Phys., 70, 1.

\ni   Beck R.,  Brandenburg A.,  Moss D.,  Shukurov A.M.,  Sokoloff D.D.,
1996, ARAA, { 34} 155

\ni Bhattacharjee A. \&  Yuan Y., 1995, ApJ 
{449} 739.

\noindent  Blackman E.G. and Field G.B., 2000, ApJ. 
{534} 984 

\ni Blackman E.G. and Field G.B. Physics of Plasmas, 2001, {8} 2407 

\ni Blackman E.G., 2002, in 
``Simulations of magnetohydrodynamic turbulence in astrophysics: recent achievements and perspectives'', 2002, eds. T. Passot and E. Falgarone, in press.

\noindent  
Brandenburg A., 2001, ApJ {550} 824

\ni  Cho J. \&  Vishniac E.T., 2000, ApJ {538} 217 

\ni  Chou, H. ApJ, 2001, { 556} 1038.

\ni   Cowling T.G., 1957,  {\sl Magnetohydrodynamics},
(Wiley Interscience, New York).

\noindent  Field G.B., \& Blackman E.G., 2002, submitted to ApJ, 
astro-ph/0111470

\ni  Frisch U.,  Pouquet A.,  L\'eorat J., and  Mazure A., 1975,
J. Fluid Mech. { 68} {769}

\ni Gruzinov, A. \& Diamond, P. H. 1994, PRL {\bf 72}, 1651.

\ni  Krause F. \&  R\"adler K.-H., 1980, {\it 
Mean-field magnetohydrodynamics and dynamo theory}, 
(Pergamon Press, New York).

\ni $^{}$    Kulsrud R.M. \&  Anderson S.W., 1992, Astrophys. J. 
{ 396} 606.

\ni   
Kulsrud R., Cen R.,  Ostriker J.P.,   and   Ryu D., 1997, 
ApJ, {480} 481

\ni  Maron J., \&  Cowley S., 2001, submitted to ApJ,
http://xxx.lanl.gov/abs/astro-ph/0111008

\ni    Maron J., and  Goldreich P. ApJ, 2001, {554} 1175.

\ni   Moffatt, H. K., 1978, {\sl Magnetic
Field Generation in Electrically Conducting Fluids}, 
(Cambridge University Press, Cambridge)

\ni   Orsag S.A., 1970, J. Fluid Mech., {41} 363.

\ni  Parker E.N., 1979, {\it Cosmical Magnetic Fields} (Oxford: Clarendon
Press).

\ni   Pouquet A., Frisch U., \&  Leorat J.,  J. Fluid Mech, 1976
{77} 321

\ni   Ruzmaikin A.A.,  
Shukurov A.M.,  Sokoloff D.D., 1988, {\sl Magnetic Fields of
Galaxies}, (Kluwer Press, Dodrecht)

\ni  Sellwood J.A., and Balbus S.A., 1999, ApJ, { 511} 660

\ni Verma M.K., 2001, submitted to Physica D, 
http://xxx.lanl.gov/abs/nlin.CD/0107069

\ni   Zeldovich  Ya. B.,  Ruzmaikin A.A., 
and  Sokoloff D.D., 1983,  {\sl Magnetic Fields in Astrophysics
}, (Gordon and Breach, New York).

\ni  Zweibel E.G. and Heiles C., 1997,  Nature, { 385} 131

\vfill
\eject

\end{document}